\documentstyle[11pt,newpasp,twoside,epsf]{article}
\def\edcomment#1{\iffalse\marginpar{\raggedright\sl#1\/}\else\relax\fi}
\reversemarginpar

\begin{document}
\title{The Characteristics of Magnetic CVs in the Period Gap }
 \author{Gaghik Tovmassian, Sergey Zharikov}
\affil{Observatorio Astron\'omico Nacional, Instituto de
              Astronom\'\i a, UNAM,
}
\author{Ronald Mennickent}
\affil{Universidad de Concepcion,  Concepcion, Chile}
\author{Jochen Greiner}
\affil{MPE, Garching, Germany}

\begin{abstract}
We have observed several magnetic cataclysmic variables located in
the range between 2 and 3 hours, known as the period gap. This
work was prompted by the recent discovery of RX\,J1554.2+2721. It
has 2.54 hours orbital period and shows almost pure cyclotron
continuum in a low luminosity state, similar to HS1023+3900,
HS0922+1333 and RBS206. These are low accretion rate polars
(LARPs) known to have mass transfer rates of order of a few
$10^{-13}$M$_{\odot}$/year. The aim of the study was to find out, if magnetic
systems filling the period gap are in any way different from their
counterparts outside that range of periods. The only significant
difference we encounter, is much higher number of asynchronous
magnetic systems to-wards longer periods than below the gap.

\end{abstract}

\section{Introduction}

Cataclysmic Variables (CVs) are close interactive binaries with
the absolute majority of their orbital periods distributed between 80
minutes and 10 hours. The bulk of CVs (Dwarf Novae and some
non-magnetic Novalikes) show a bimodal distribution of orbital
periods with a well pronounced deficiency of systems between 2
and 3 hours, known as the Period Gap. For a long time it's been argued
and now commonly accepted that the magnetic systems does not
follow that pattern and the distribution of magnetic CVs does not show
such bi-modality (Webbink \& Wickramasinghe 2002). The explanation
that they suggest is based on magnetic breaking models that allow
narrowing of the period gap for sufficiently high magnetic moment
of the primary star immediately after which the mass transfer is
driven by gravitational waves. The model does a good job of
describing the orbital period distribution of magnetic CVs, but it
envisages much higher rates of mass transfer than were observed in
a few magnetic CVs in the period gap and around it. We were able
to conduct spectrophotometry of a sample of mCVs in the period gap
to compare their characteristics with others. We also observed one
object that lies above the period gap but also shows an extremely low
mass transfer rate.

The question is whether the mass transfer decrease is directly
associated with the period gap, and, thus detection of magnetic CVs
there is a matter of selection effect (primarily X-ray emission),
or if it is an intrinsic characteristic of particular systems.

\begin{figure}
\plotfiddle{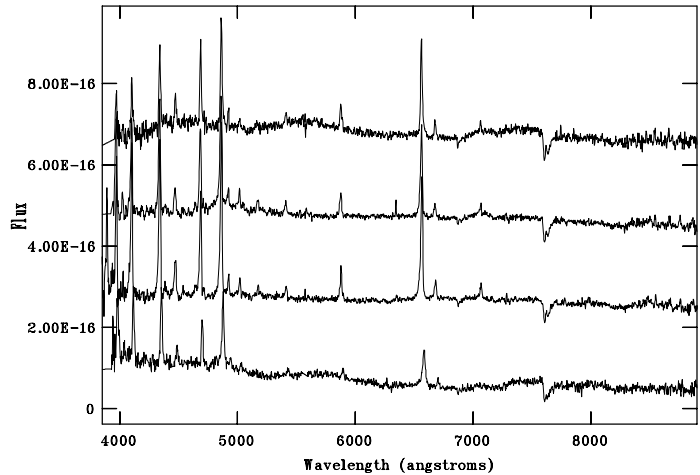}{52mm}{0}{65}{65}{-200}{-310}
\caption{ Spectra of V381\,Vel\, in different phases}
\end{figure}

\section{Observations and basic characteristics of the selected objects}

There are 51 systems in between 2 \& 3 hours in the living edition
of the Downes (1997/2002) catalog of Cataclysmic Variables. 25 of them are
proven to be magnetic systems. We  collected low resolution
spectrophotometric data on 3 of them. Although they constitute
only a small fraction of the systems inside the period gap, they
exhibit wide range of features characteristic of Polars. We also
observed HS0922+1333 which is remarkable for its low mass accretion
rate, its pure cyclotron spectrum and being located above the Gap.

Observations were conducted mainly at the 2.1m telescope of OAN in
Mexico. One object (V381Vel) was observed from La Silla with 3.5m
NTT (for details of observations see Tovmassian et al. (2003a,
2003b).

\subsection{V381 Vel = RX J1016.9-4103}

\begin{figure}
\plottwo{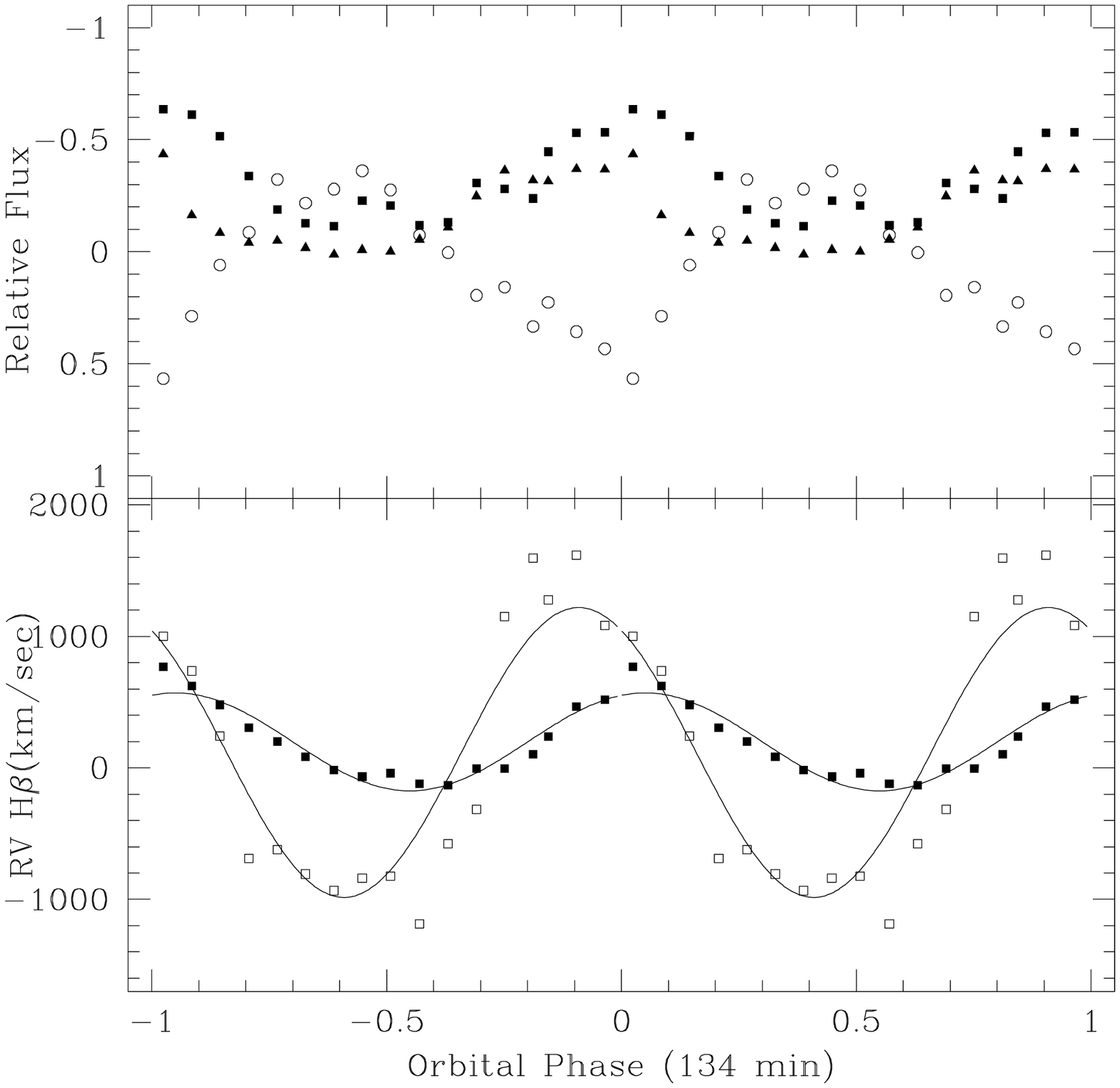}{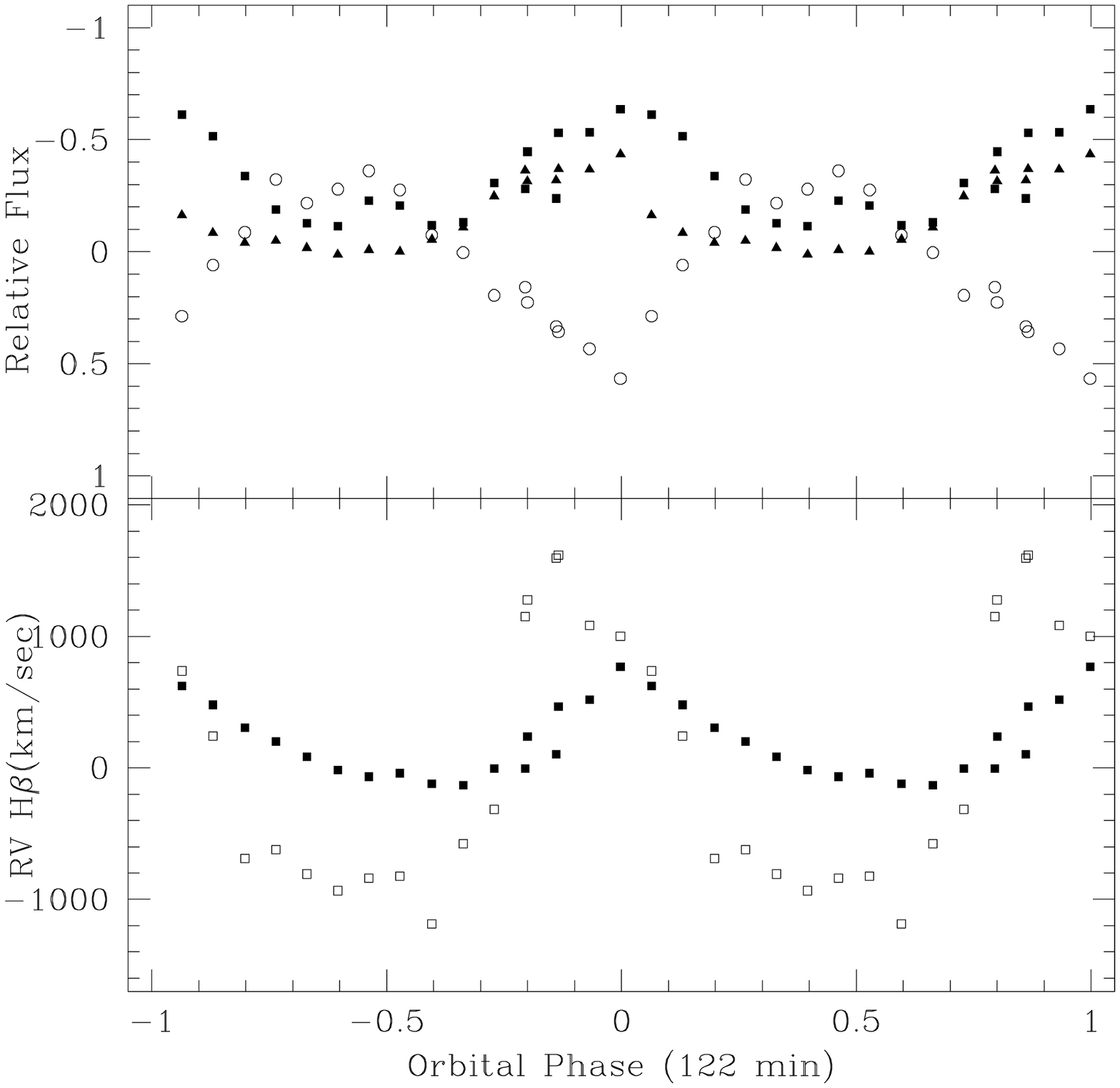}
\caption{Radial velocities of H$\alpha$ line and fluxes of cyclotron lines
and continuum of V381\,Vel\, 
folded with orbital (left) and spin (right) periods. }
\end{figure}

Optical counterpart of this Polar was identified by Greiner \&
Schwarz (1998). They estimated P$_{\rm orb}$=134 min. 
The system has  one pole of B=52 MG or 41 MG. Vennes et al. (1999) 
later cited a shorter P$_{\rm orb}$=122 min, without
elaborating much on the other properties.
We observed one full period in April 1999 without being aware
of the controversy. Complete orbital coverage allowed us to
conclude that the system actually has two poles. At phases
0.20-0.25 we can see mix of both poles, as Greiner \& Schwarz
(1998) did. At phase 0.5 no cyclotron humps are apparent. Later at
$\phi=0.7$ the 2nd harmonic of 41 MG pole dominates. Finally at
phase 0.95 through 0.1 the stronger 52 MG pole is seen (see Fig
1).
The emission lines were successfully separated into two components.
None of these components arises from the irradiated secondary
star. Rather they are emitted from different parts of the
accretion stream. Folding the data with both announced periods and
fitting a $sin$ curve strongly favors 134 min as an orbital period
of the system (Fig 2, lower panels). 
\begin{figure}[t]
\plotfiddle{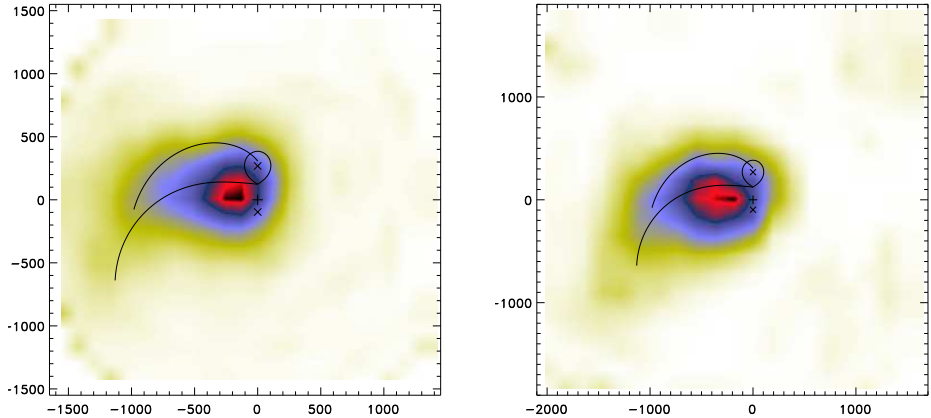}{52mm}{0}{70}{70}{-220}{-145}
\caption{Doppler maps  of V381\,Vel. H$\alpha$ in the left panel and He\,{\sc ii} in the right.}
\end{figure}
Next, we integrated the flux in  three different bands. They were
selected to reflect the flux variation in the continuum relatively free of
emission lines (including cyclotron lines), and two other narrow
bands centered on the cyclotron features. 
The folding of the continuum fluxes, presumably from the
self-eclipsing WD, with the orbital (134 min) period is poor (Fig
2, upper left panel). The same data folded with 122 min period
greatly improves the light curve, but distorts the radial velocity
curve (right panel, note the overlapping of data points at $\phi$ 0.8-0.9).
We believe that this Polar is not synchronized and that P$_{\rm
orb}=134$ min, while 122 min is the spin period of the WD. The
difference is  about 10\%, which in unprecedent for Polars.
The orbital phasing can be independently confirmed by the Doppler
tomograms. The very characteristic pattern of the mass transfer flow
is oriented precisely, as one would expect with the right parameters
and  phasing (Fig 3). Once again the absence of an irradiated secondary in
the V381\,Vel should be stressed.

\subsection{RX J1554.2+2721}

\begin{figure}[t]
\plotone{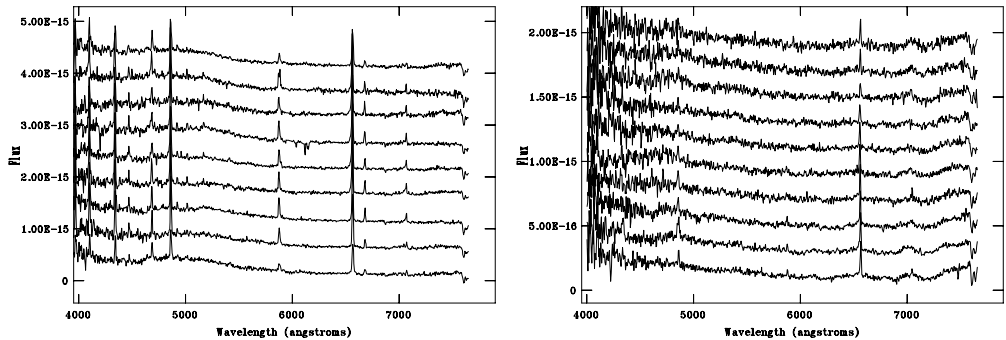}
\caption{Spectra of RXJ1554 in high (left) and low (right) states around the orbital period.}
\end{figure}
This object was recently identified as a Polar (Tovmassian et al.
2001).  We estimated P$_{\rm orb}=2.753$ hours. Shortly after, Thorstensen
\& Fenton (2002) refined it to 2.53 hours, placing it exactly in
the center of the period gap. Its main features include frequent
switches from the high to the low luminosity states, extremely low
mass transfer rate in the low state and a striking similarity to
a pair of LARPs (HS09, HS10 and RBS 206). Actually, the latter
conclusion was based upon a single spectrum obtained in the low
luminosity state. We re-observed RXJ1554 until we caught it in the
low state again. New observations confirmed that the object goes from the high
state to the low within months. We got a full orbital coverage of
the system in the high in the low state. We confirmed the estimate of the magnetic 
field strength of order of $B=30$ MG. The object probably has a weaker second pole. 
In the Fig 4 we present samples of spectra of RX\,J1554\,
at different phases in the low and high states. 
\begin{figure}
\plotfiddle{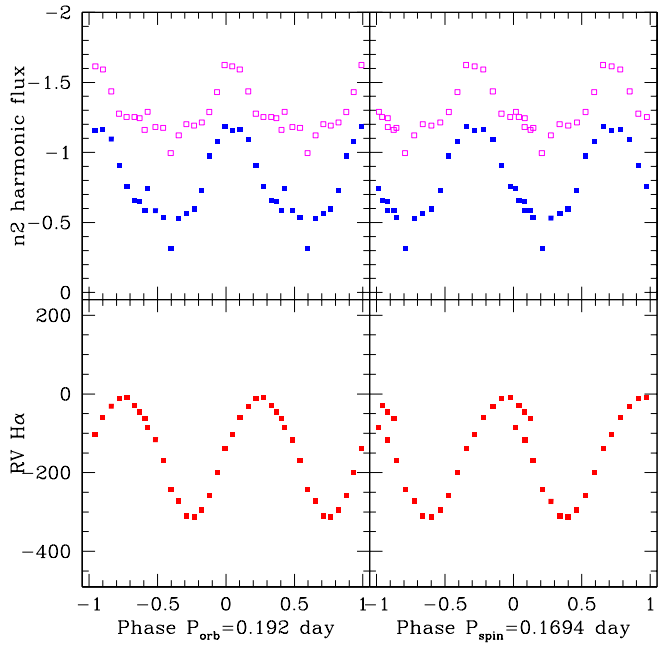}{84mm}{0}{45}{45}{-150}{-75}
\caption{ Radial velocities of H$\alpha$ line and fluxes of cyclotron lines
of HS09 folded with orbital (left) and spin (right) periods.}
\end{figure}

Here again we separated the emission lines into two
components. In this case however, the dominant component arises
from the irradiated secondary star, an M4 red dwarf. This
component is basically the only line emission  observed in the low state. The
stream is not detected or is extremely faint. Doppler maps in 
Tovmassian et al. (2001) and Thorstensen \& Fenton (2002) 
testify to  the presence of a strongly irradiated secondary and a weak
stream contribution unlike the V381\,Vel.
The system seems  to be synchronized.  At least no deviations are
detected at our precision level. 
These are preliminary results and we expect to obtain more useful
information from wealth of  data that was acquired.

\subsection{HS1023+3900}

We observed HS10 along with RXJ 1554 in January \& April 2002 primarily to
check if it undergoes states of higher luminosity. In both
occasions the object looked exactly as described in Reimers, Hagen
\& Hopp (1999). There is  a continuing evidence of an irradiated
secondary (H$\alpha$ in emission), as was reported earlier by Schwope
(2001).

\subsection{HS0922+1333}

In our small survey we also included HS09 (Reimers \& Hagen 2000),
because, it is so far the only system with similar cyclotron
spectra and extremely low mass transfer rate, which is firmly
located outside the 2-3 hour period range, above it. Our main goal
as in the case of HS10 was to check if it shows high luminosity
state episodes. Results so far are negative. However, an analysis
of its orbital parameters yielded a nice surprise. 
We were able to measure the orbital period directly from the H$\alpha$ 
emission line and the Na\,{sc i} absorption doublet.

The radial velocity measurements of H$\alpha$ in emission as well as of
Na{\sc i} absorption doublet folded with period reported
by Reimers \& Hagen (2000) showed obvious discontinuity. 
Although our coverage was barely enough
to overlap periods and it was not possible to estimate the period
precisely, the quality of data was good enough to unambiguously
establish that the orbital period exceeds the spin period by at
least 30 minutes and is of order of 4.6 hours. 
The narrow band flux measurements  centered on the
cyclotron emission lines and folded with both, the newly determined orbital, 
and the previously reported 4.07 hour periods confirm that the latter is a 
better fit and most probably represents the spin period of the primary.
In the Fig 5 similar to Fig 2 the radial velocity and flux measurements 
are folded with the proposed orbital and spin periods. The overlapping of phases
occurs around $\phi\, 0.1-0.2$.    
The lack of synchronization was suspected by the 
discoverers of the object.

\begin{figure}
\plotfiddle{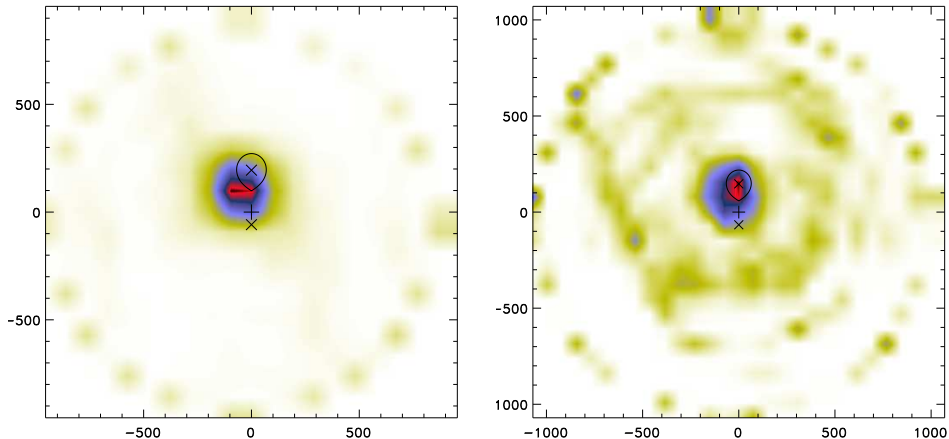}{53mm}{0}{70}{70}{-250}{-155}
\caption{ Doppler maps of HS09. H$\alpha$ in the left panel, Na\,{\sc i} in the right}
\end{figure}

Doppler tomograms of H$\alpha$ (left) and Na{\sc i} (right) 
lines demonstrate the presence of an irradiated secondary star in the system and
also some evidence of a mass transfer stream immediately near the L$_1$
point. This is an interesting observation taking into account 
the meager mass transfer rate of $3\times10^{-13}$M$_{\odot}$/year.

\section{Conclusions}

We were able to demonstrate that Polars located in the period gap show all
range of spectrophotometric features proper to mCVs: 

RXJ1016 was observed only in the
high state. It shows a significant contribution from the accreting
matter flow in  emission lines and no secondary irradiation. There 
is no evidence of significantly low mass transfer rate in this case. 
It is probably an asynchronous rotator. 

RXJ1554 was frequently observed switching from
high to low states. In the high state it shows bluer continuum, strong
emission lines both from the irradiated secondary and the stream.
In the low state it becomes similar to the LARPs. Probably
synchronous. It bridges the seemingly large differences between
LARPS (Schwope 2001) and ordinary Polars.

\begin{figure}
\plotfiddle{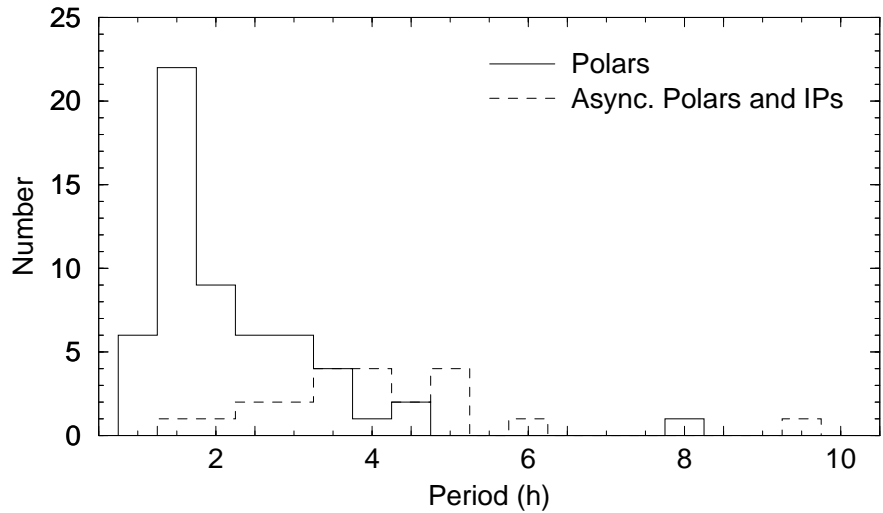}{54mm}{0}{73}{73}{-210}{-50}
\caption{Orbital period distribution of magnetic CVs}
\end{figure}

HS10 has never been observed in high state. The secondary
is irradiated. Probably synchronous. 

In the low states the last two objects  show extremely low accretion
rates (~$10^{-13}$ M$_{\odot}$/year), but so does HS09  with a period of  4.6 hours
and RBS 206, that may be below the period gap  at 90 min. Thus, we may speak 
about existence of a distinct class of Polars with a sufficiently lower
mass transfer rates than originally known. However there is no strong evidence,
linking low mass transfer rate with the ``classical'' period gap. 

The considered objects have magnetic field strengths and secondaries whose spectral
classes are  within the usual range of values.

Redrawing the orbital distribution graph  with the new assignments 
(HS09 \& RXJ1016 as asynchronous polars) creates an impression that 
the 2-3 hour range is a ``melting point'' where magnetic systems reach synchronization.
This conclusion is in a good agreement with the  prediction drawn by Webbink 
\& Wickramasinghe (2002) from their models (although they expect at least $10^{-10}$ M$_{\odot}$/year).
LARPs are still understudied objects and they can provide important insights to the nature of mCVs.

Discovery of a couple of highly asynchronous polars arises the question of the 
definition of Intermediate Polars. Are they simply magnetic CVs
with a spin period of the primary sufficiently different from the  orbital period or 
is the strength of magnetic field the most important defining parameter?

\end{document}